\documentclass[conference,10pt]{IEEEtran}

\usepackage[top=0.75in, bottom=1.03in, left=0.625in, right=0.625in]{geometry}
\usepackage{epsfig,rotating,setspace,latexsym,amsmath,epsf,amssymb,amsfonts,bm,theorem,subfigure,epstopdf}

\usepackage{cite}
\usepackage{bbm}
\usepackage{enumerate} 
\usepackage{layout}

\newcommand\FTD{\textcolor{black}}

\newcommand{\BK}{\textcolor{black}}

\usepackage[ruled,vlined,linesnumbered]{algorithm2e}
\usepackage{subcaption} 
\usepackage{multirow}
\usepackage{listings}
\usepackage{makecell}
\usepackage{graphics}
\usepackage{array}

\newcolumntype{?}{!{\vrule width 1pt}}

\usepackage{color,booktabs,multirow}

\IEEEoverridecommandlockouts
\allowdisplaybreaks
\bstctlcite{IEEEexample:BSTcontrol}
\begin{document}

	\title{Deep Reinforcement Learning Based Routing for Heterogeneous Multi-Hop Wireless Networks }
 
	\author{\IEEEauthorblockN{Brian Kim, Justin H. Kong, Terrence J. Moore and Fikadu T. Dagefu}
 \IEEEauthorblockA{U.S. Army DEVCOM Army Research Laboratory, Adelphi, MD, 20783, USA}

 }

\maketitle


\begin{abstract}
Routing in multi-hop wireless networks is a complex problem, especially in heterogeneous networks where multiple wireless communication technologies coexist. Reinforcement learning (RL) methods, such as Q-learning, have been introduced for decentralized routing by allowing nodes to make decisions based on local observations. However, Q-learning suffers from scalability issues and poor generalization due to the difficulty in managing the Q-table in large or dynamic network topologies, especially in heterogeneous networks (HetNets) with diverse channel characteristics. Thus, in this paper, we propose a novel deep Q-network (DQN)-based routing framework for heterogeneous multi-hop wireless networks to maximize the end-to-end rate of the route by improving scalability and adaptability, where each node uses a deep neural network (DNN) to estimate the Q-values and jointly select the next-hop relay and a communication technology for transmission. To achieve better performance with the DNN, selecting which nodes to exchange information is critical, as it not only defines the state and action spaces but also determines the input to the DNN. To this end, we propose neighbor node selection strategies based on channel gain and rate between nodes rather than a simple distance-based approach for an improved set of states and actions for DQN-based routing. During training, the model experiences diverse network topologies to ensure generalization and robustness, and simulation results show that the proposed neighbor node selection outperforms simple distance-based selection. Further, we observe that the DQN-based approach outperforms various benchmark schemes and performs comparably to the optimal approach.  




\end{abstract}

\section{Introduction}

Routing in multi-hop wireless networks is a complex problem that involves establishing a route from the source to the destination while meeting specific network performance requirements, such as end-to-end rate and latency. One of the most well-known traditional methods for routing is the open shortest path first protocol using Dijkstra's algorithm \cite{dijkstra1959note} to find the shortest path to the destination by weighting the edges of the graph based on link performance among nodes. However, this approach lacks adaptability to network dynamics, and the overhead of collecting all necessary information in advance is substantial as the network size grows, forcing the routing algorithm to be decentralized for multi-hop wireless networks. 

To overcome these shortcomings, decentralized algorithms for multi-hop wireless networks have been studied considering different network objectives and environment \cite{kandelusy2020cognitive,senanayake2018decentralized}, where each node chooses the next node by evaluating multiple conflicting factors, making the routing problem inherently a sequential decision-making problem. Here, each decision depends on the current state of the network, which contains only the information from close relay nodes, making the algorithm decentralized. Therefore, such routing can be defined as a Markov decision process (MDP) \cite{bellman1957markovian}, which makes it an ideal candidate to be solved using reinforcement learning (RL) based approaches. 

A type of RL called Q-learning, a model-free value-based RL, has been used to find the optimal route for multi-hop networks \cite{chang2004mobilized,arafat2021q}, where each node stores a table of Q-values that is used to select the next node with the highest Q-value. Here, the Q-values are iteratively learned and improved over time by taking correct actions that give a high reward. The use of Q-learning for the routing of multi-hop networks has been extended to heterogeneous networks (HetNets) in \cite{kong2024decentralized,kim2024failure} where HetNets use multiple communication technologies to increase network throughput, latency, and covertness. Note that different communication technologies have different standards, protocols, and operate in different parts of the spectrum. However, these studies considering Q-learning for routing face scalability and generalization challenges where each node needs to store a large table of Q-values for a given network that needs to be changed when the network conditions change.

  

Thus, a deep RL (DRL) approach has been introduced to handle routing decisions in a more robust and scalable manner using deep neural networks (DNNs). Unlike traditional RL, DRL-based routing methods are not limited to a predefined topology, but instead dynamically adapt to network conditions such as node mobility and change of topology by experiencing numerous environments during training. Moreover, DRL techniques can incorporate multiple quality-of-service factors, including latency and bandwidth, into the routing decision-making process \cite{zhao2020deep, rathore2019deep}. One of the most prominent DRL techniques is the deep Q-Network (DQN), which leverages DNNs as the Q-value function by enabling nodes to learn optimal policies through experience replay and target networks \cite{mnih2015human}. This reduces the need to store an extensive table with Q-values at each node and allows the model to generalize across different network topologies, making routing more efficient even in dynamic and large-scale multi-hop networks. 
In~\cite{cui2021scalable}, the DQN was applied to wireless routing to maximize the end-to-end rate by mitigating the interference from other transmissions.
In this work, each node employs a DNN to determine the next hop and the spectrum to use for transmission, based on information from the neighbor nodes selected based on proximity, as a distance-based channel model is assumed, which ensures higher channel gains. DRL using graph neural networks (GNNs) for wireless networks has been investigated in \cite{zhang2024decentralized} to maximize end-to-end rate by jointly optimizing the route and transmit power at each node. While these DRL methods demonstrate strong performance across various topologies, they rely on a distance-based channel model, which limits their applicability in real-world scenarios where fading effects arise due to the specific environmental geometry and material properties. Further, with the growing demand for HetNets to enhance coverage and rate, it remains unclear how DRL can effectively establish a route by leveraging the multiple communication technologies \BK{with different bandwidths.}


Therefore, in this paper, we consider a routing for a heterogeneous multi-hop wireless network using DQN, where each node aims to maximize the end-to-end rate by selecting one of the available communication technologies \FTD{(as well as available subbands for each communication technology)} and the next hop among the neighbor nodes. Here, we consider different channel characteristics for different communication technologies due to their different operating frequencies. Furthermore, we propose different methods for selecting neighbor nodes to exchange information in heterogeneous multi-hop wireless networks, as choosing the physically closest nodes is not always optimal when multiple communication technologies are available and channel quality \FTD{is strongly correlated with the local fading effects which are frequency-dependent.}
The key contributions of the paper are: 
\begin{itemize}
    \item To the best of our knowledge, this is the first work exploring routing in heterogeneous multi-hop wireless networks using DQN, where each node is equipped with multiple communication technologies featuring distinct channel characteristics. A DNN is employed at each node to select both the next hop and the communication technology \FTD{(and corresponding subbands)} used to transmit, aiming to maximize the end-to-end rate of the route. 
    \item We propose two approaches for selecting neighbor nodes to exchange information that is fed into the DNN, such as channel-based and available rate-based neighbor node selection, and compare them to traditional distance-based selection. Note that different neighbor node selection leads to different state and action spaces, significantly affecting the DNN performance.
    
    \item To enhance the generalizability and robustness of our DNN, we train with a variety of network topologies. Then, the performance of the proposed DRL-based routing approach is compared to multiple benchmark methods, including the widest path-based algorithm. 
\end{itemize}

\FTD{\section{System Model and Problem Formulation}}
\label{sec:SystemModel}


\FTD{\subsection{Network Model}}
\label{subsec:NetworkModel} 

We consider a HetNet consisting of a source, a destination, and $N$ potential relay nodes a subset of which will constitute the route from source to destination. All nodes, including source and destination, are equipped with $M$ different communication technologies and each communication technology $m$ has $B^{(m)}$ available subbands \BK{where the bandwidth of the communication technology is fixed as $\Omega^{(m)}$ and divided equally among the $B^{(m)}$ available subbands, i.e., $\Omega^{(m)}/B^{(m)}$. Here, each communication technology uses a unique frequency band and has its own channel characteristics.}


We first define $\mathcal{G}$ as the set of all possible routes that can be formed. Then, the possible route is defined as 
\begin{equation}
    \mathbf{G} = \{e_0,e_1, e_2, \cdots,e_{N_{G}},e_{N_{G}+1} \} \in \mathcal{G},
\end{equation} 
where $e_{i} \in \mathcal{E}$ is the $i$th relay node in the \BK{set of all nodes $\mathcal{E}$}, 
$N_{G}$ is the number of relay nodes, $e_0$ is the source node, and $e_{N_{G}+1}$ is the destination node. Further, a possible set of communication technologies \BK{and subbands} used for the route is defined as 
\begin{equation}
    \mathbf{T} = \{t(0),t(1),t(2), \cdots,t({N_{G}}) \},
\end{equation}
where $t(i)\in \mathcal{T}$ is the communication technology and corresponding subband used at node $e_{i}$ and $\mathcal{T}$ is the set of all communication resources (combination of all available communication technology and subbands) available, defined as
\begin{equation}
    \mathcal{T} = \{t^{(1)}_1,t^{(1)}_2,t^{(1)}_3, \cdots, t^{(1)}_{B^{1}}, \cdots, t^{(M)}_{B^{M}-1}, t^{(M)}_{B^{M}} \},
\end{equation}
with $|\mathcal{T}| = \sum_{i=1}^{M} B^{(i)}$. Here, $t_{j}^{(i)}$ represents the communication technology $i$ using subband $j$.


We assume that each node uses only one communication resource for transmission and prohibits it from using the same resource for transmission and reception.
Further, we restrict the flow from visiting the same node more than once to avoid forming a loop in the route. Specifically, for the node $e_i$,
\begin{equation}
     t(i) \neq t({i-1}), \;\;\; e_{i} \ne e_{j}, \;\forall j \ne i.
\end{equation}

\subsection{Performance Metric}
When the transmitter node $e_i \in \mathbf{G} \setminus \{e_{N_{G}+1}\}$ in the route $\mathbf{G}$ selects communication resource $t(i)$ from communication technology $m$ and sends $x \sim \mathbb{C}(0,1)$, the received signal at the receiver $e_{i+1}$ is
\begin{align}\label{eq:Received_Receiver}
    y_{e_i,e_{i+1}}^{t(i)} = &\sqrt{P} h_{e_i,e_{i+1}}^{t(i)} x+ 
    +\underbrace{\sum_{k \ne i}\sqrt{P} h_{e_k,e_{i+1}}^{t(i)} x}_{\text{interference}}  + n_{e_i,e_{i+1}}^{t(i)},
\end{align}
where $P$ is the transmit power at node $e_i$ and $h_{\alpha,\beta}^{t(i)}$ denote the channel from the transmitter $\alpha$ to the receiver $\beta$ for resource $t(i)$. Further, $n_{e_i,e_{i+1}}^{t(i)}[l] \sim \mathcal{CN}(0,\Omega^{(m)}N_0/B^{(m)})$ is additive white Gaussian noise (AWGN) at the receiver, where $N_0$ is noise power spectral density. Note that we consider interference that comes from other hops established within the flow. Thus, the signal-to-interference-plus-noise ratio (SINR) at the receiver $e_{i+1}$ for resource $t(i)$ is defined as 
\begin{align}\label{eq:SINR}
\text{SINR}^{t(i)}_{e_{i+1}} = \frac{P | h_{e_{i},e_{i+1}}^{t(i)} |^2}{\Omega^{(m)}N_0/B^{(m)} + \sum\limits_{k \ne i}P |h_{e_k,e_{i+1}}^{t(i)}|^2 },
\end{align}
and the rate of a single-hop link between nodes $e_i$ and $e_{i+1}$ is
\begin{align} \label{eq:Rate_link}
	R^{t(i)}_{e_{i},e_{i+1}} = \frac{\Omega^{(m)}}{B^{(m)}}\log_2\big(1+\text{SINR}&^{t(i)}_{e_{i+1}}\big).
\end{align}
Since the lowest rate of the single-hop links in the route determines the overall rate, the end-to-end rate of route $\mathbf{G}$ using $\mathbf{T}$  set of communication resources is defined as
\begin{align} \label{eq:Rate_route}
	&R(\mathbf{G},\mathbf{T}) =  \underset{ e_i \in \mathbf{G} \setminus \{e_{N_G+1}\}}{\min} R^{t(i)}_{e_{i},e_{i+1}}.
\end{align}


\subsection{Optimization Problem Formulation}
In this paper, we aim to maximize the end-to-end rate by determining the set of actions, such as selecting the node to transmit and the communication resource (communication technology and subband) for the transmission. Mathematically, the optimization is formulated as  
\begin{align} \label{eq:objective function}
	\underset{ \mathbf{G} \in \mathcal{G}, \mathbf{T} \in\mathcal{T}}{\max} & \;\; R(\mathbf{G},\mathbf{T}) \\ 
	 \text{s.t.} \;\;\;\;&\;\; t(i) \neq t(i-1), \;\;\;\; \forall i \in\{1,\ldots,N_G\}\nonumber\\
     & \;\; e_{i} \ne e_{j}, \;\;\;\;\;\;\;\;\;\;\;\;\;\;\;\forall j \ne i, \forall i \in\{1,\ldots,N_G\}.\nonumber
\end{align} 
Note that the optimization problem (\ref{eq:objective function}) is an integer programming problem, which is an NP-hard problem with a large number of discrete optimization variables involved. \FTD{Furthermore}, the problem is highly complex due to the interdependency between hops caused by interference in equation (\ref{eq:Received_Receiver}). Specifically, if two hops use the same communication resource, then the transmitted signal of a hop interferes with the other hop, making spectrum allocation and communication resource selection crucial for the routing problem. 

To tackle this difficulty, we model it as an MDP, since the routing problem involves sequential decision-making in a dynamic environment. Here, each node observes local information from neighbor nodes without the help of a centralized server and decides the action in a distributed manner. This makes a partially observable Markov decision process (POMDP) which motivates the use of multi-agent RL where each node acts as an agent. Furthermore, unique characteristics of the wireless environment due to the wireless medium, such as dynamic channel conditions and interference, motivate the need to come up with a DRL agent.

\section{Deep Reinforcement Learning Agent for Routing} 

\FTD{In this work, we propose a DRL-based approach to solve the optimization problem (\ref{eq:objective function}) by exploiting a single DRL agent that is shared by all nodes along the flow.} \BK{Specifically, we deploy a single, pre-trained DRL agent across all nodes, as opposed to different DRL agents being trained at different nodes, to significantly reduce the number of parameters to train and improve scalability and generalizability due to the agent being trained with diverse flows and network topologies.} 


Further, we aim to train our DRL agent to work across all communication resources. A simple way to do this is to train a DRL agent by aggregating features for all communication resources and neighbor nodes, and deciding the optimal next node and communication resource. However, this approach significantly increases the state and action spaces for the DRL agent resulting in slower convergence. Therefore, in this paper, we use our DRL agent for one communication resource at a time using information only from neighbor nodes, then select the best node and corresponding communication resource with the highest estimated Q-value. By doing so, the agent has fewer actions and state space to consider, which helps the agent to converge faster to the optimal decision. Note that the same DRL agent is used for all communication resources.


\subsection{Action Space}
First, we define the action space for the DRL agent for one communication resource $t$. 
Then, the set of all possible actions at the frontier node $e$, the current node that is making the decision as in Fig. \ref{fig:states}, is represented as
\begin{align*} 
	\mathcal{A}_{e}^{(t)} = \{ a_{e,\eta_1}^{(t)}, a_{e,\eta_2}^{(t)},\cdots, a_{e,\eta_{N_{e}}}^{(t)}   \},
\end{align*}
where $a_{e,\eta_i}^{(t)}$ is the action that frontier node $e$ transmits data to $i$th neighbor node $\eta_i$ using resource $t$ and $N_{e}$ is the number of neighbor nodes. Note that the number of neighbor nodes $N_{e}$ and how to select $(\eta_1,\cdots, \eta_{N_{e}})$ significantly impacts the performance of the DRL agent, as it alters both the size and composition of the action space.


\subsection{Proposed Neighbor Node Selection}\label{sec:node selection}
We propose two different approaches to select the neighbor nodes, such as channel-based and rate-based approaches, and compare them with the distance-based approach where the frontier node selects $N_{e}$ neighbor nodes based on distance.
\subsubsection{Channel-based \FTD{neighbor node selection}}
For this approach, the frontier node $e$ computes the average channel gain across all communication technologies with each neighbor node $k$ as $\frac{1}{|\mathcal{M}|}\sum_{i=1}^{|\mathcal{M}|}|h^{t_1^{(i)}}_{e,k}|$
. Then, the node selects $N_e$ number of nodes with the highest average channel gain as the neighbor nodes where the features of these nodes will be aggregated and input to the DRL agent to estimate the Q-value for each action.


\subsubsection{Rate-based \FTD{neighbor node selection}}
Similarly, the frontier node $e$ calculates the rate $\frac{1}{|\mathcal{M}|}\sum_{i=1}^{|\mathcal{M}|}R^{t_1^{(i)}}_{e,k}$ for node $k$. Then, it selects $N_e$ number of nodes with the highest rate as the neighbor nodes where the features of these nodes will be aggregated and input to the DRL agent to decide an action. Note that this approach considers not only the channel but also interference from other hops and available bandwidth.

\subsection{ State Space}
\begin{figure}[t]
    \vspace{0.02in}
    \centering  \includegraphics[scale=0.125]{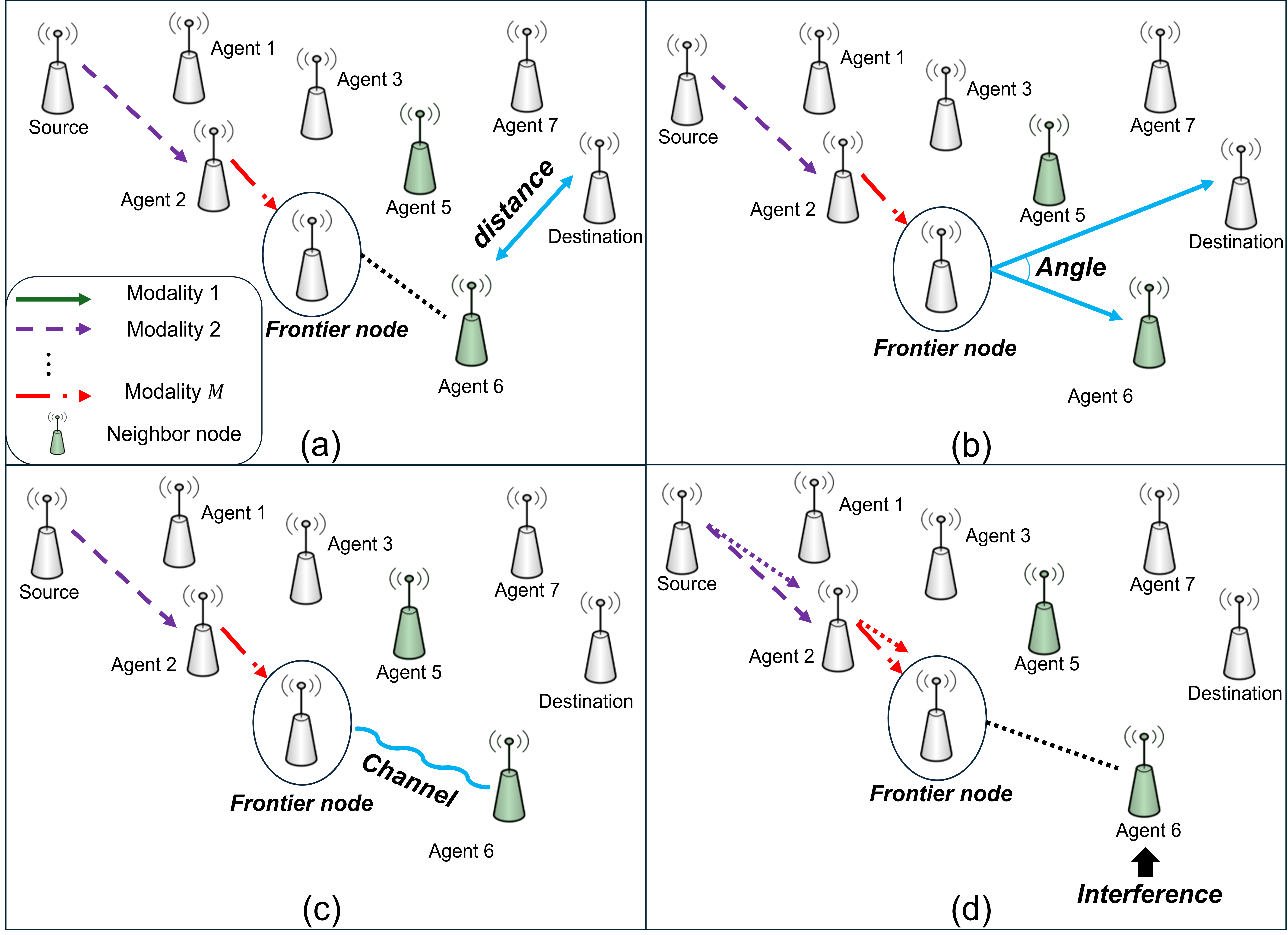}
    \caption{States that DRL agent at the frontier node gathers such as (a) distance, (b) angle, (c) channel gain, and (d) interference.}
    \label{fig:states}

\end{figure}

The state space $\mathcal{S}$ of the DRL agent for one communication resource should be defined to represent crucial factors of $N_{e}$ neighbor nodes for action decision. Since the goal of our DRL agent is to maximize the end-to-end rate of the route, we consider the following as the elements of the state space, as seen in Fig. \ref{fig:states}.
\begin{enumerate}[(a)]
     \item Distance between the neighbor node and the destination.
     \item Angle difference between the frontier node to the destination and the frontier node to the neighbor node.
    \item Channel gain between the neighbor node and the frontier node.
    \item Total interference the neighbor node is experiencing in the current communication resource.
\end{enumerate}

Here, we have geometric information, such as (a) and (b), which indicates how close \FTD{and in what direction } the neighbor node is to the destination. Also, we have wireless link information, such as (c) and (d), representing the link quality between the frontier node and the neighbor node. As a result of defining the four-feature state space for each neighbor node, the total state space used to decide the action for the frontier node at a specific communication resource is $4N_{e}$ factors.

Note that information such as coordinates, cumulative interference, and channel gain can be readily exchanged between nodes via the control and non-payload communications (CNPC) link, as described in \cite{cui2021scalable, zhang2024decentralized}. These types of data are commonly used in routing and resource allocation, with information exchange typically limited to local neighbor nodes. Thus, the DRL agent incurs minimal additional overhead compared to existing algorithms when acquiring local features.

\subsection{Reward Function}
In this paper, we use DQN to approximate the Q-value function by leveraging DNNs. Here, the Q-value, estimating the value of a state-action pair, is used to decide the action at the agent by selecting the action with the highest Q-value. Conventionally, Q-value $Q(s_t, a_t)$ is updated using the immediate reward and the future reward as follows.
\begin{align}
    Q(s_t, a_t) = Q(s_t, a_t) + \alpha [ r_t + \gamma \max_{a'} &Q(s_{t+1}, a') \nonumber\\
    &- Q(s_t, a_t) ],
\end{align}
where $r_t$ is the immediate reward at time $t$, $Q(s_{t+1}, a')$ is the future reward from the next state, $\alpha$ is the learning rate, $\gamma$ is the discount factor, and $(s_t,a_t)$ is the action $a_t$ state $s_t$ pair.

Therefore, to calculate the Q-value, we first need to develop a reward function. Since the goal of the agent is to maximize the end-to-end rate of the route, it is natural to set the reward as the bottleneck rate of the established route. However, setting the bottleneck rate as a reward hinders the agent from finding the optimal solution since the earlier hops can dominate the bottleneck rate, meaning the immediate and the future rewards are set as the rate from the earlier hops, making the action chosen at the current step irrelevant. As a consequence, the agent is unable to learn during training.

To solve this issue, we define the reward function with only the future reward where the future reward is defined as the bottleneck rate after the route has been established, similar to the approach done in \cite{cui2021scalable}. Specifically, during training, the agent observes the state $s_t$ and the action $a_t$ until it reaches the destination. Then, after the route is fully established, state-action pairs along the route are stored with the future reward, which is the bottleneck rate of the established route, as follows. 
\begin{equation}\label{eq:Q-value}
    Q(s_t, a_t) = \underset{ e_i \in \mathbf{G} \setminus \{e_{N_G+1}\}}{\min} R^{t(i)}_{e_{i},e_{i+1}}.
\end{equation}


\subsection{Deep Q-Network}\label{sec:DQN} 
\begin{figure}[t]
    \vspace{0.02in}
    \centering  \includegraphics[scale=0.1]{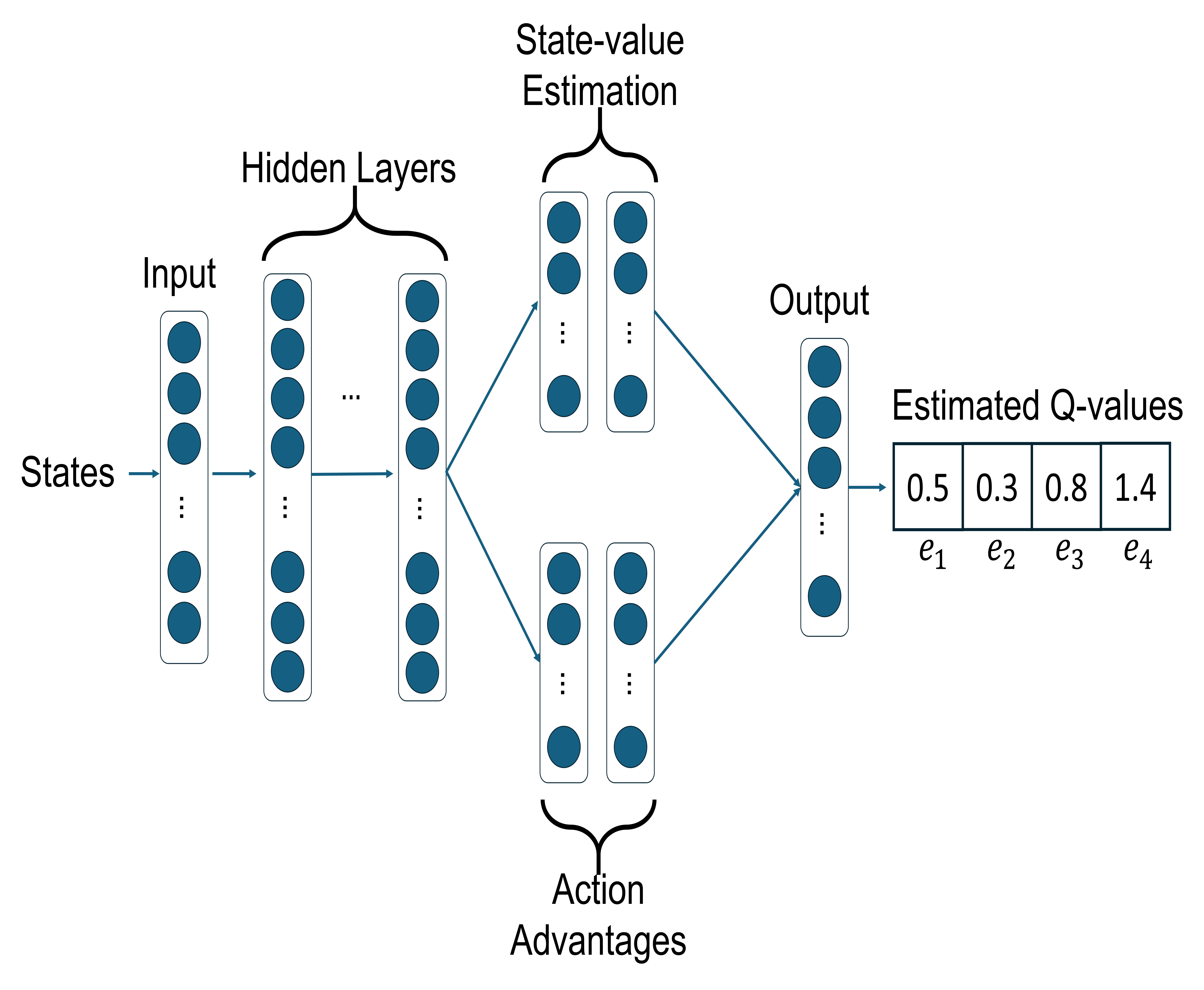}
    \caption{Dueling-DQN architecture with the input being the states and the output being the estimated Q-values for communication resource $t$.}
    \label{fig:dqn}
\end{figure}

Due to the state features being real-valued variables and the inherent dynamics of wireless networks, the agent needs to be able to generalize well to the unseen environment with distinct state features. For this purpose, we exploit DQN where the DNN is used to predict the Q-value given the state-action pairs. During training, we store the state input as a vector with the length of $4N_e$ and the corresponding Q-value for each action which is Equation (\ref{eq:Q-value}). \FTD{We build on the dueling-DQN architecture} from \cite{wang2016dueling} for our agent, which consists of two parts, one estimating the state values and the other estimating the action preference, as seen in Fig. \ref{fig:dqn}. 

During the training of DQN, we use the experience-replay method studied in \cite{mnih2015human} where the $\epsilon$-greedy policy is used to gather experiences for the replay buffer. The $\epsilon$-greedy policy \cite{sutton2018reinforcement} is a policy where the node $e$ takes an action based on the Q-value (exploitation) from the DNN with probability $1-\epsilon$ and chooses one random action from the action space (exploration) with probability $\epsilon$ to explore new actions that have not yet been considered. As a result of the $\epsilon$-greedy policy, we obtain state-action pairs where the corresponding bottleneck rate is calculated after the route is fully established. Then, we store (state, action, reward) tuples in the replay buffer which will be used for DQN training. Note that since the state and action are used during training, how to select them, as previously mentioned in Section \ref{sec:node selection}, greatly affects the DQN performance. To ensure convergence of the DQN, we set $\epsilon=0$ at the end of the training and let the agent collect and train without any exploration. After the training, we use DQN for each communication resource and decide the action at the frontier node that gives us the highest Q-value by comparing all estimated Q-values for all communication resources.   

\section{Simulation Results}
\subsection{Simulation Environment}
\begin{figure}[t]
    \centering  \includegraphics[scale=0.28]{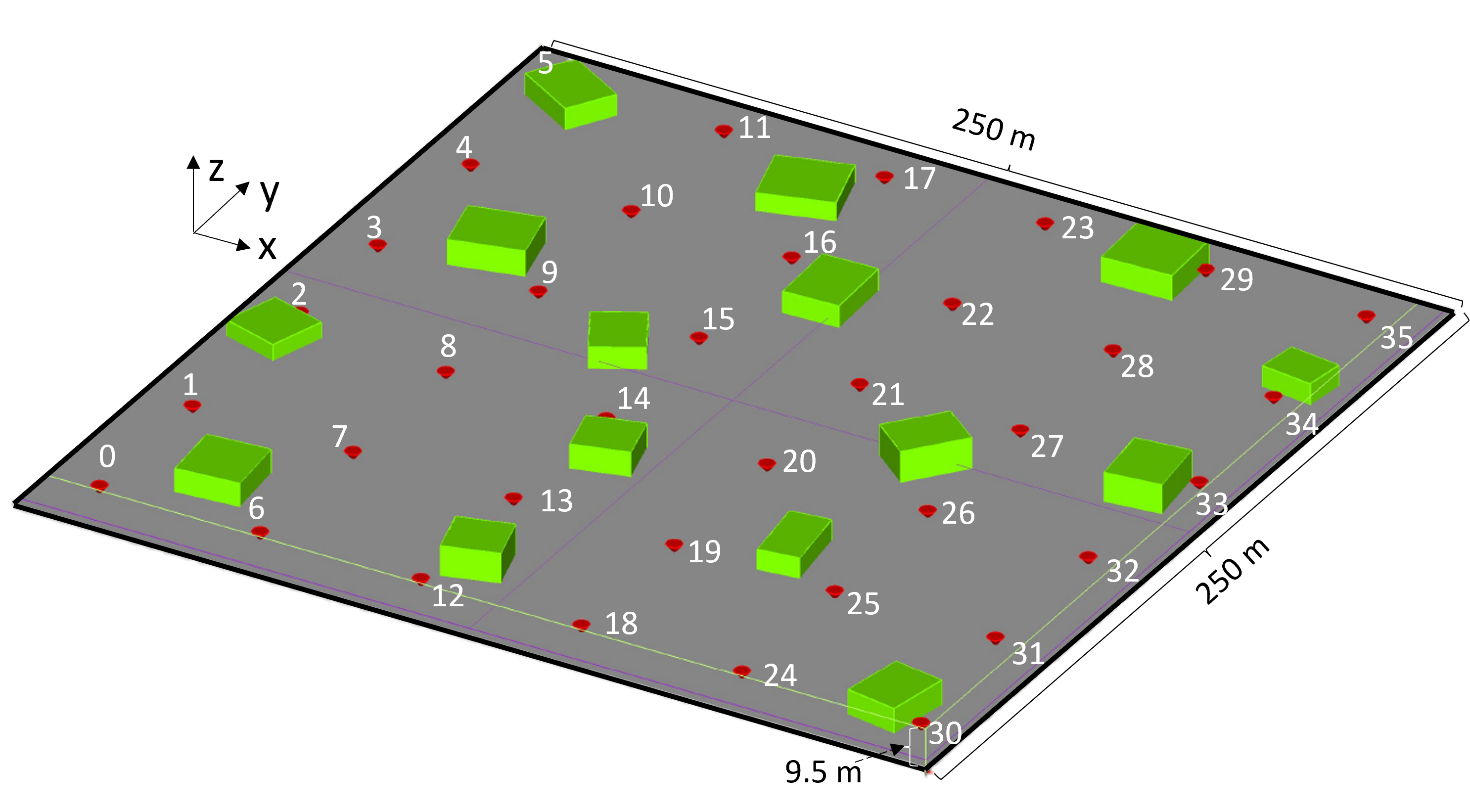}
    \caption{3D simulation environment with 36 nodes.}
    \label{fig:simulation}
\end{figure}

We consider a simulation environment consisting of a 3D multi-building environment (250$\times$250$\times$9.5) m$^3$  with concrete buildings (green boxes) and 36 legitimate nodes (red dots) as in Fig. \ref{fig:simulation}. Each legitimate node is equipped with the same three different wireless communication technologies with center frequencies $f_{c,m_1} = 400$~MHz, $f_{c,m_2}= 900$~MHz, and $f_{c,m_3}= 2.4$~GHz. The bandwidth allocated to a subband of technology $m_i$ is $0.01 f_{c,m_i}/B^{(m_i)}$~MHz. Note that the total bandwidth of each communication technology is set to 1\% of its center frequency, which provides a realistic representation of practical scenarios. Also, we consider different number of subbands $B^{(m_i)}$ during simulations. The channel data for all three technologies, all 36 nodes, was computed on a grid of densely sampled receiver locations using a high fidelity ray tracing based approach \cite{EMCUBE}. To create dynamic network topologies, we randomly select 31 nodes out of 34 nodes for each episode during training and testing, where the location of the source and destination are fixed to \textit{node 0} and \textit{node 35} from Fig. \ref{fig:simulation}. We train our DRL agent using 1 million episodes and test the performance with 1000 different topologies. 


As mentioned in Section \ref{sec:DQN}, we consider dueling-DQN architecture consists of input layer of size $4N_e$, 3 hidden layers with 300 nodes, two hidden layers with 300 and 150 nodes for each layer for the state-value estimation part, two hidden layers with 300 and 150 nodes for each layer for the action advantages part, and output layer of size $N_e$.

\begin{algorithm}[t]

\DontPrintSemicolon
\SetAlgoNoEnd
\SetAlgoLined
Define source $S$ and destination $D$ nodes.\\
Define vertices (nodes) $V$ for given network topology. \\
Define frontier node $F = S$.\\
\While{$F \ne D$}{
Create a graph $(V,E)$ where edges $E$ are defined as $e_{i,j} = \max_{t(i)} R^{t(i)}_{i,j}$.\\
Obtain a route $[F, N_1, N_2, \cdots, D]$ from the widest path algorithm.\\
Update $F = N_1$.
}
Route = All intermediate frontier nodes $F$.
\caption{Widest path-based algorithm \label{alg:widest path}} 
 \end{algorithm}

\subsection{Benchmark Schemes for Routing}
Due to the unavailability of the optimal solution to the multi-hop wireless routing considering interference, we consider various benchmark schemes as follows.
\begin{itemize}
    \item \textit{Strongest neighbor}: Select the neighbor with the strongest average wireless channel from the frontier node.
    \item \textit{Best direction to destination}: Select the neighbor with the best direction such that the angle difference between the agent-to-neighbor and the agent-to-destination is smallest.
    \item \textit{Closest to destination}: Select the neighbor closest to the destination node. 
    \item \textit{Least interfered}: Select the neighbor and communication resource with the least interference.
    \item \textit{Largest data rate}: Select the neighbor and communication resource with the highest link capacity.
    \item \textit{Destination directly}: Select destination directly from the source.
    \item \textit{Widest path-based algorithm}: Obtain a route from the frontier node to the destination using the widest path algorithm. Then, the frontier node selects the next hop and communication resource from that route considering the interference from already selected transmitting nodes. A detailed algorithm is presented in Algorithm \ref{alg:widest path}. 
\end{itemize}
Note that for \textit{Strongest neighbor}, \textit{Best direction to destination}, \textit{Closest to destination}, and \textit{Destination directly} schemes, the communication resource is selected with the highest rate between the frontier node and the selected neighbor node.


\begin{figure}[t]
    \vspace{-0.2in}
    \centering  \includegraphics[scale=0.43]{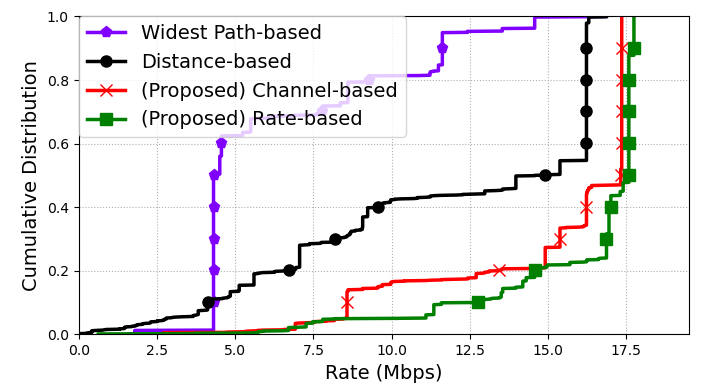}
    \caption{Comparing different node selection approaches for the DRL agent when $N_e =5$ and $B^{(m_i)} = 5$ for all $i$.} 
    \label{fig:node selection}
\end{figure}

\subsection{Routing Performance of DRL Agent}

We begin by analyzing how different strategies for selecting neighbor nodes impact the routing performance, as illustrated in Fig. \ref{fig:node selection}. In the simulation, we set $N_e =5$ and $B^{(m_i)} = 5$ for all $i$. The results show that the widest-path-based approach performs poorly compared to the DRL agent utilizing various neighbor node selection methods. Among different neighbor node selection approaches, the distance-based selection yields the weakest performance, as it often fails to select suitable candidates for data transmission. In contrast, the rate-based neighbor node selection outperforms the others, as it considers not only channel quality but also interference and available bandwidth $\Omega^{(m)}/B^{(m)}$.

\begin{figure}[t]
    \centering  \includegraphics[scale=0.41]{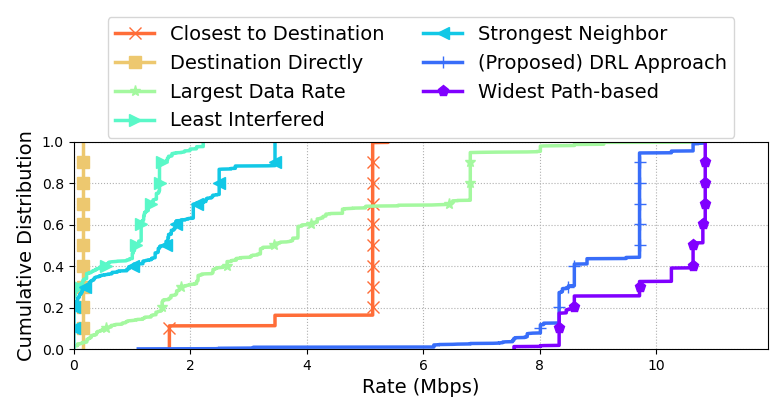}
    \caption{Comparing DRL approach with various benchmark schemes when $N_e =10$ and $B^{(m_i)} = 15$ for all $i$.}
    \label{fig:15subband}
\end{figure}

Different benchmark schemes are compared to the DRL agent with rate-based neighbor node selection in Fig. \ref{fig:15subband}. Here, we consider $N_e =10$ and $B^{(m_i)} = 15$ for all $i$. It is seen that the widest path-based approach outperforms other benchmark schemes, including the DRL approach, where it achieves 10 Mbps on average. This is due to the ability to get the optimal route when the interference effect can be removed. Specifically, the widest path-based scheme inherently disregards the interference from future hops as seen in Fig. \ref{fig:node selection}. However, if there are enough subbands to mitigate the interference, the widest path-based scheme, a centralized approach, becomes the optimal approach. Therefore, this result shows that the performance of the DRL agent is comparable to the optimal centralized approach, where it achieves 9.01 Mbps on average.

\begin{figure}[t]
    \centering  \includegraphics[scale=0.43]{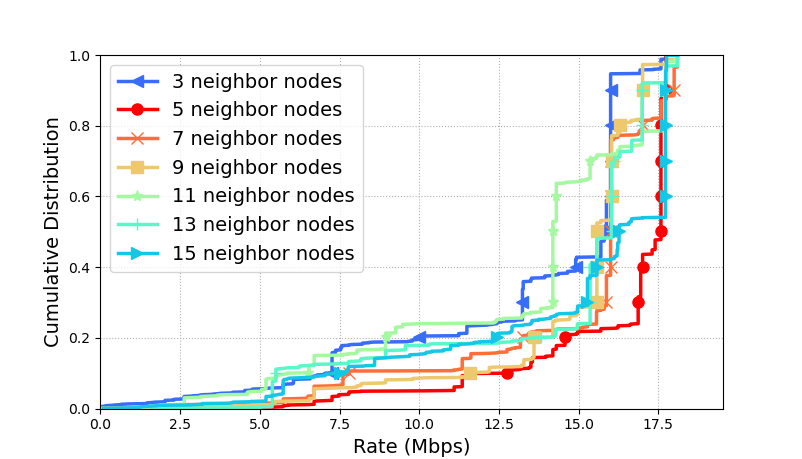}
    \caption{Comparing the performance of DRL for different numbers of neighbor nodes $N_e$ when $B^{(m_i)} = 5$ for all $i$.}
    \label{fig: neighbor node num}
\end{figure}

Now, we vary the number of neighbor nodes $N_e$ and observe how the performance of the DRL agent changes. Here, we consider the rate-based neighbor node selection and 5 subbands for each communication technology. It is observed in Fig. \ref{fig: neighbor node num} that the $N_e = 5$ case outperforms other cases. This is due to the increase in action and state spaces for the higher number of neighbor nodes, which leads to a decrease in performance due to the slower convergence with a limited number of episodes. Also, it is observed that the $N_e = 3$ case performs worse than the $N_e = 5$ case, even though it has smaller action and state spaces, since the number of neighbor nodes is small and fails to include the essential nodes. Thus, it is crucial to properly select the size of the neighbor nodes because it not only determines the performance of the DRL agent but also the communication overhead due to the messages that are being sent to get information.

\section{Conclusion} \label{sec:Conclusion}
This paper presented a DQN-based routing framework for heterogeneous multi-hop wireless networks, where each node selects both the next-hop relay and the communication resource to maximize end-to-end rate. By incorporating realistic channels and exploring different neighbor selection strategies beyond a simple distance-based approach, the proposed methods improved routing performance and adaptability. Simulation results showed that training across diverse network topologies helped the robustness and generalization of the model and outperformed other benchmark schemes. Furthermore, our model even performs comparably to the widest path-based approach when a large number of subbands are available, a scenario in which the widest path-based method becomes optimal due to the elimination of interference from other hops.




\bibliographystyle{ieeetr}
\bibliography{references}

\end{document}